\documentclass[sigconf,screen]{acmart}
\usepackage[utf8]{inputenc}
\usepackage{graphicx}
\usepackage{appendix}
\usepackage{multicol}
\usepackage{ragged2e}
\usepackage{array}

\title{A Post a Day Keeps the Doctor Away: Sharing Personal Information on Self-Diagnosis Platforms}
\author{Roopa Bhat}
\affiliation{%
  \institution{Columbia University}
  \city{New York}
  \country{U.S.A.}
}
\email{rsb2178@columbia.edu }

\author{Lord Crawford}
\affiliation{%
  \institution{Columbia University}
  \city{New York}
  \country{U.S.A.}
}
\email{lord.crawford@columbia.edu}

\author{Nicole Hong}
\affiliation{%
  \institution{Columbia University}
  \city{New York}
  \country{U.S.A.}
}
\email{nhh2112@columbia.edu}
\date{April 2021}

 \setcopyright{acmcopyright}
 \copyrightyear{2021}
 \acmYear{2021}
 \acmDOI{10.XXXX/XXXXXXX}

 \acmConference[HCI Seminar]{}{Spring 2021}{Columbia University, NY}
 \acmBooktitle{HCI Seminar, Spring 2021, Columbia University, NY}
 \acmISBN{978-1-4503-XXXX-X/18/06}

\begin{document}

\begin{abstract}
    For many, it can be intimidating or even impossible to seek professional medical help if they have symptoms of an illness. As such, some people approach platforms like Reddit or Quora for a community-based conversation in an attempt to diagnose themselves. In this paper, we unearth the factors affecting people’s willingness to share personal information with online self-diagnosis platforms and forums. From an online survey and in-depth interviews, we present who this population of users are, and what, where, and why they are posting. Our evaluation finds that tech-savvy young adults are more likely to post on online platforms about potentially sensitive or highly specific topics for convenience, fast response, and a sense of community. We found that anonymity, distrust of physicians, and prior experience with platforms largely played a role in people’s willingness to share personal health information online for self-diagnosis purposes.
\end{abstract}

\maketitle

\section{Introduction}

In the last decade, the Internet has enabled a space for interactive health information-seeking \cite{pewresearch, chi2020connections}, thereby helping countless individuals diagnose themselves with medical conditions \cite{mccarthy2017did, schoenherr2014interactions}. A 2013 nationwide study from the Pew Research Center showed that a third of Americans went online to figure out a medical condition \cite{pewresearch}. This emergence of health information-seeking can be linked to health anxiety, otherwise known as cyberchondria, which refers to “the unfounded escalation of concerns about common symptomatology” \cite{white2009cyberchondria}. 

Some people have turned to online health communities in hopes of receiving support from others who have medical expertise or share a similar experience. This implies that people are very willing to disclose personal health information to receive advice on their symptoms \cite{de2014mental}. An example of this is largely seen on Reddit--a social news application consisting of user-generated discussion forums and polls--where people can post their health inquiries and have a wide network potentially offer guidance or support. In this context, users can also secure anonymity when posting through the use of 'throwaway accounts’ \cite{de2014mental}. Since anonymity and data privacy are great concerns when sharing personal health information, and more generally information online \cite{leon2013matters}, Reddit serves as an optimal place for health communities.
This paper takes a step further from the existing literature, focusing on online platforms and forums, such as Reddit and Quora, to examine the population of people willing to divulge personal health information in their self-diagnosis journey. To better understand this space, we tackled four related research questions:
\begin{enumerate}
    \item Who is this population of posters (people willing to share personal health information on online platforms and forums)?
    \item Where on the internet is this population posting? 
    \item What is this population posting? 
    \item Why is this population posting? What are their motivators?
\end{enumerate}

After conducting an initial online survey followed by in-depth interviews, we found some compelling takeaways from this population of ‘posters,’ namely how anonymity impacts user behavior on online health communities, the cases of sharing personal information on forums, and how this phenomenon is granting people access to a better understanding of their health. 

% Sample code to add figure in one column
% \begin{figure}[h]
%     \centering
%     \includegraphics[width=0.95\linewidth,keepaspectratio]{table1.png}
%     \caption{Sample.}
%     \label{fig:fig1}
% \end{figure}

\section{Related Work}
\textit{2.1 Online Health Information Seeking}

Previous research informs us of how vast the field of online health information seeking is, including the various lengths people have gone to self-diagnose online \cite{pewresearch} and the mediums established to help them do so. The following seem to have inspired the emergence of online health information-seekers: health anxiety, online health communities, and the ease of web search \cite{chi2020connections, mccarthy2017did, schoenherr2014interactions, white2009cyberchondria}. A study conducted by Leiden University’s Department Public Health \& Primary Care displayed various Web-search strategies in health information seeking \cite{kwakernaak2019patients}, specifically showing the volume of symptom-based approaches and decisions participants made in verifying their hypotheses (self-diagnosis).\\ \\
\textit{2.2 Personal Health Information Sharing}

An empirical study conducted by Yuting Liao of University of Maryland tackled what the motivators are for people to share personal health information on social media \cite{liao2019sharing}. It was concluded that people care about spreading awareness for specific medical conditions, showing solidarity, and providing social support to their followers if they share a similar experience. These are all characteristics we see manifested in spaces on Reddit. Notably, within mental health subreddits, there is a greater amount of personal health information disclosure due to anonymity and a strong sense of solidarity among users \cite{de2014mental}. We focus more broadly on why people are gravitating towards online forums like Reddit to self-diagnosis. 

This work investigates the relationship between willingness to share personal health information when self-diagnosing online across multiple web platforms. Furthermore, it is not restricted by a particular domain or health topic. While existing literature explores online health information-seeking, this work focuses on the factors that affect diagnosis-seeking behavior online.

\section{Methods}
\textit{3.1 Survey}

We used Google Forms to create an online survey that examined 1) a participant’s general use with online self-diagnosis platforms and forums, 2) concerns when sharing personal information on the internet and 3) in-depth experiences with posting self-diagnosis queries on online platforms. In total, we had 21 participants complete the survey. We recruited participants mainly by utilizing Reddit’s subreddits (see Appendix): forums or sub-communities dedicated to a specific topic, where individuals can post comments, articles, media, and vote on content they find compelling. We also recruited from our personal networks. All participants read and signed an informed consent form before participating.\newline\newline
\textit{3.2 Interviews}

We reached out to those in our personal network, who consented to being interviewed and sharing their age and gender, to learn more about their experience with online self-diagnosis platforms. The interviewees were split into two groups: posters (those who have actually made personal posts) and observers (those who have interacted with and visited these platforms, but have not made personal posts). Out of the five interviewees, two were posters and three were observers. The posters were asked about their personal characteristics, the content of their post, the platform they posted on, and the reasons and motivations behind their post. The observers were primarily asked about their reasons for not posting. 

\section{Results}
\textit{4.1 Online Survey Findings} 

Through our survey (n=21), we found that 100\% of participants have used websites to search for symptoms. However, only a third of the participants took an additional step to make a personal post on an online platform specific to their situation. In this paper, we separate those who have interacted with these platforms, but have never made a personal post (observers) and those who have posted (posters).

% Code for a sample table - full width
\begin{table*}[h]
    \centering
    \begin{tabular}{|>{\raggedright\arraybackslash}p{2.5cm} |>{\raggedright\arraybackslash}p{7cm}|>{\raggedright\arraybackslash}p{7cm}|}
         \hline 
         \textbf{} & \textbf{Observers} & \textbf{Posters}\\
         \hline
       \textbf{Considerations for sharing personal information online} & Almost 40\% cited data privacy as a key concern when sharing personal information online \newline \newline More than half of participants consider the potential of being stalked or the credibility of the platform before sharing & More than 70\% cited anonymity as a key concern when sharing personal information online \newline \newline Top considerations before sharing personal information online were related to anonymity and whether they could be identified by those who know them\\
        \hline
         \textbf{General social media activity} & More than 20\% of non-posting participants do not consider themselves as active posters on social media platforms \newline\newline Less than half of the participants who are active users of social media post every day & 60\% of participants who consider themselves active on social media post on social media every day \newline\newline Less than 30\% consider themselves as inactive users on social media platforms\\
         \hline
         \textbf{General usage of online platforms to search up symptoms} & WebMD is the most popular platform when searching up symptoms, followed by MayoClinic and Reddit forums equally & WebMD is the most popular platform when searching up symptoms, followed by MayoClinic and Reddit forums equally \newline\newline 100\% of personal posts were made on Reddit forums\\
         \hline
    \end{tabular}
    \caption{Characters of Observers vs. Posters. Observers are participants who do not have experience making a personal post. Posters are participants who have at least one experience making a personal post.}
    \label{tab:table1}
\end{table*}

It is interesting to note that observers were primarily concerned with data privacy while posters were primarily concerned with anonymity. The top reason for not posting was that personal queries were not necessary, as the answers from general searches were sufficient (see Table 1). 

The main attraction point for seeking medical advice online was the fast response rate and convenience (see Figure 1). Notably, those who have posted before on self-diagnosis platforms sought medical advice online because of the difficulty with traveling in-person, distrust of physicians, lack of health insurance, and prior success with online health platforms, more than those who have not posted before. One participant explained: \textit{“I prefer to be in control of my medical experience. A wrong diagnosis by several [doctors] when I was 15 left me in horrible pain as I suffered with hydronephrosis of my left kidney until I was 23. A nurse discovered it was my kidney by accident. From this experience I learned that [doctors] are human and make mistakes. I must advocate for myself and be as educated as possible about my condition(s)."}

% Sample code to add figure in one column
\begin{figure}[h]
     \centering
     \includegraphics[width=1\linewidth,keepaspectratio]{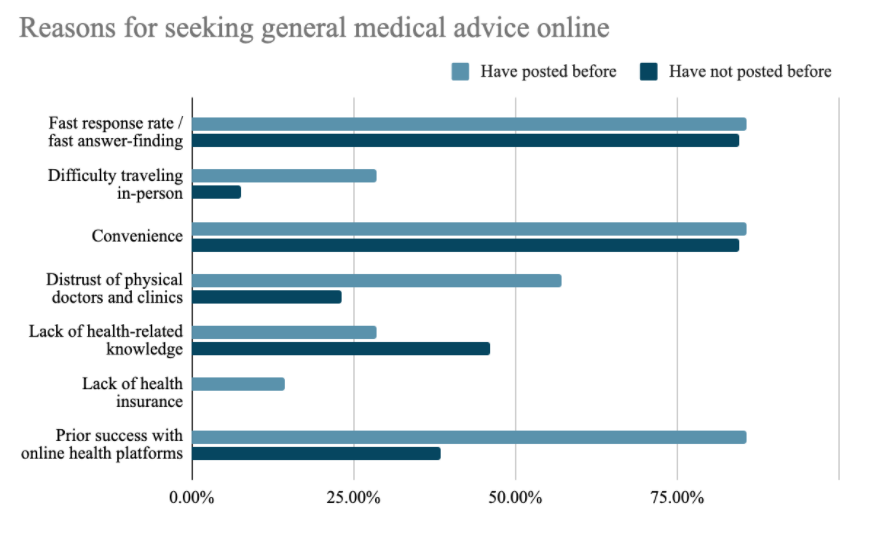}
     \caption{Reasons for seeking general medical advice online between observers and posters (n=21)}
     \label{fig:fig1}
 \end{figure}
Focusing on posters, we report the results of their posting experience. The reasons for making a personal post lined up similarly with their reasons for seeking general medical advice online (see Figure 2), with the need for a quick response being the most popular followed by prior success with posts and distrust of physicians. Additionally, almost half of the participants who have had experience with personal posts created a post regarding mental health. 

% Sample code to add figure in one column
\begin{figure}[h]
     \centering
     \includegraphics[width=1\linewidth,keepaspectratio]{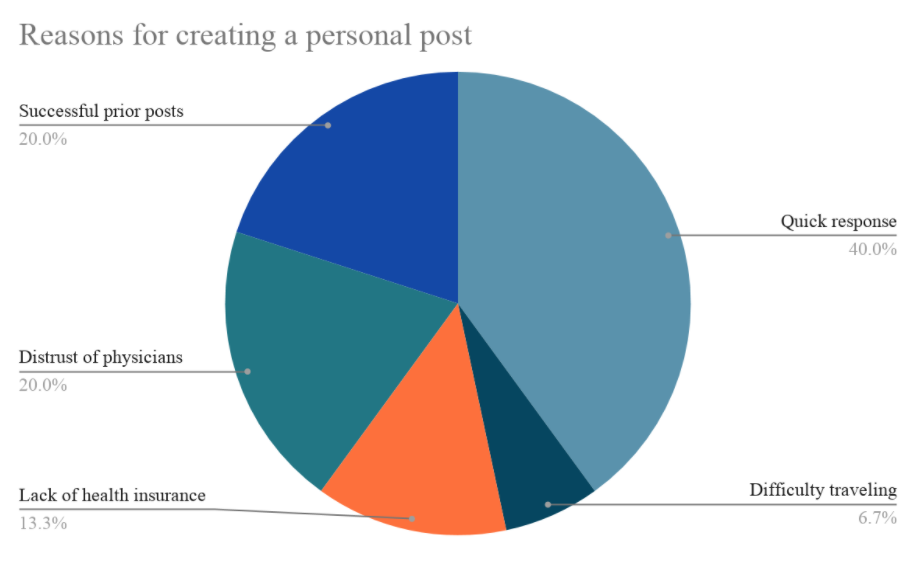}
     \caption{Reasons for creating a personal post on online platforms for self-diagnosis purposes (n=21)}
     \label{fig:fig2}
 \end{figure}
Finally, we found that all posters shared only non-identifiable information: gender, age, and a non-identifiable photo. None of the posters were comfortable with sharing an identifiable photo and 67\% of the participants who responded to the question, \textit{“To what extent were you comfortable with sharing the following information in your post?”}, were not comfortable with sharing their name. These results relate well to the fact that more than 70\% of the posters cited anonymity as a primary concern.  

\section{Discussion}
Our research question explored the who, what, where, and why of people’s usage of self-diagnosis platforms and forums. We now discuss the implications of our findings. \newline\newline
\textbf{Who - Characteristics of Posters}

There may be a potential link between comfortability with technology and posting medical queries on online platforms. Our interviews indicate that basic technological literacy can be a prerequisite for making a personal post, while other times, it can also be a disqualifier. For example, those who feel comfortable discerning credible information already on the internet may not deem it necessary to make a personal post, as indicated by Observer A:\textit{ “I’ve never run into a situation where I’m looking for something so rare/specific that I cannot find anything anywhere."} On the other hand, Poster A explains that his ability to discern factual versus inaccurate information on the Internet makes him more comfortable posting on platforms such as Reddit, because he is confident in his ability to gauge the credibility of any answer he may receive. Thus, technological literacy can impact whether someone deems it necessary to make a personal medical post.

Users who are more active on social media--defined as posting at least for significant life events--are also more likely to create a personal post on online platforms. This makes sense as activity on social media can be indicative of less hesitation regarding sharing personal information with a larger online network. That said, although Observer A is pretty active on social media, she \textit{“doesn’t think social media is the place to share her ailments and health issues.”} Part of the reason may be the content that is being shared; she is comfortable liking and commenting on posts online, but any public engagement beyond that is intimidating: \textit{“If I make a personal post, I would have to give away information about myself that I wouldn't be comfortable sharing.”} This indicates that posting on social media is viewed differently than posting on online self-diagnosis platforms. 
\newline\newline
\textbf{What - Content in Personal Posts}

Participants who post typically asked questions about highly specific or potentially sensitive topics that may be difficult to talk about in person, like illicit drugs and mental health. One poster quoted that they were searching the consequences of \textit{“taking 5 hpt 10 hours after MDMA,”} an illegal hallucinogenic drug taken mostly at clubs or raves. The poster may have resorted to online forums to avoid potential legal consequences. This becomes a robust implication when coupled with the finding that all posters only posted non-identifiable information, like their age, gender, or a non-identifiable photo. Poster A explained his decision to post about skin conditions: \textit{“I posted specifically about my skin condition because it’s not as clear--how do you explain a picture? I did search through the 'skin conditions' flair on the subreddits but these are difficult to compare to your own, given different lighting, different skin colors etc. When I couldn’t find something similar, I decided to make a personal post.”} It can be deduced that conditions that are not easily explained by a list of clearly defined symptoms, such as skin conditions or mental health conditions, were more likely to appear in novel posts because they were difficult to search in existing information. \newline\newline
\textbf{Where - Frequented Online Platforms}

It was interesting to find that while WebMD was the most popular place for posters and observers to search for symptoms online, all the personal posts were made on Reddit. One reason may be that the existing information is too general, pushing posters to ask specific queries. Secondly, similar to the role of technological literacy and credibility of sources in participants' likelihood of creating a personal post, Reddit was a popular choice due to anonymity, prior success with the platform, and comfortability with the platform (due to high usage). Poster B explains that she gravitated towards the anonymity of Reddit because she did not want to ask questions about sexual health to her family and friends. Furthermore, credibility played a role in deciding which platform to utilize. Poster B indicated that Reddit provided more credible answers and a more intuitive user interface than Yahoo Answers, a platform she had used in the past to answer her medical questions. On the other hand, while Observer C observed that people made personal comments on different sites, he was hesitant to post because \textit{“[the websites] lacked credibility.”} Lastly, Poster A exemplified how frequent usage of the platform led to comfortability with making a post: \textit{“I’m a frequent Reddit user so I knew there would be a subreddit for pretty much anything in this world.”}\newline\newline
\textbf{Why - Reasons for Sharing Personal Health Information}

Finally, we delve into why posters share personal information for self-diagnosis purposes. The general consensus seemed to be that convenience and a fast response rate were the main reasons behind utilizing self-diagnosis platforms. Poster A decided to make a post because it was more convenient than going to see a doctor: \textit{“I just wanted an answer right away. I assumed I would have to go to a dermatologist, which means I would have to go to my PCP and then get a referral, which would all be a lot of effort.”} Some people also utilize these platforms because they want to be able to talk about sensitive subjects without being identified. For mental health issues, online self-diagnosis platforms provide an open space wherein people can share their detailed experiences, knowing that they will receive validation and comfort from others. Furthermore, as indicated by Poster B, for topics like sexual health, younger posters can receive accurate guidance without having to inform their parents; Reddit provided an \textit{“adult audience”} that could \textit{“guide [her] in the right direction.”} She sought a community, since her environment was not conducive to conversations surrounding mental health issues. 

It was noteworthy that some interviewees were not searching for health information for themselves, but for others. Observer B utilized online platforms to look up information about her father’s cancer and was willing to make a post if necessary. This was a key benefit of online platforms; medical information could be shared and obtained for someone else even if the patient was not interested in receiving medical attention. However, Observer B also noted that she preferred an in-person diagnosis because she knew the doctors were required to be confidential with the information and her health information would not be \textit{“floating around the internet.}” \textit{“I don’t want unsolicited advice from random people,”} she explained. On the other hand, 20\% of posters created a personal post due to the distrust of physicians. Being able to learn about other “patients” online in detailed ways seemed to be a motivator for some, while a discouraging factor for others.
\section{Limitations and Future Work}

Both the survey and interviews shed light on the various factors influencing a person’s decision to post on self-diagnosis platforms and forums. However, we encountered limitations in the types of questions we could ask due to privacy concerns surrounding personal information, like demographics. In the future, we would like to increase our sample size to make more representative and robust generalizations regarding factors that affect people’s willingness to share personal health information online.

Our study also revealed interesting topics that could motivate future research. For example, one interviewee highlighted that many people have taken to Instagram or the like to share that they have received the Covid-19 vaccine, indicating that people are sharing personal medical information online for non-diagnostic purposes. Future work may entail looking at the various factors influencing a person’s decision to share personal medical information outside of self-diagnosis purposes. Additionally, research on analyzing the efficacy of self-diagnosis via online platforms may be an interesting and important next step. 

\section{Conclusion}

Through an empirical study consisting of an online survey and in-depth interviews, this paper demonstrates that tech-savvy young adults are more likely to post on online platforms about potentially sensitive or highly specific topics for convenience, a fast response, and a sense of community. We found that anonymity, distrust of physicians, and prior success with platforms played a role in people’s willingness to share personal health information online for self-diagnosis purposes. All in all, our findings provide a first look into the importance of self-diagnosis platforms for personal health. 

\bibliographystyle{ACM-Reference-Format}

\begin{thebibliography}{9}

%%% the Web address.

\ifx \showCODEN    \undefined \def \showCODEN     #1{\unskip}     \fi
\ifx \showDOI      \undefined \def \showDOI       #1{#1}\fi
\ifx \showISBNx    \undefined \def \showISBNx     #1{\unskip}     \fi
\ifx \showISBNxiii \undefined \def \showISBNxiii  #1{\unskip}     \fi
\ifx \showISSN     \undefined \def \showISSN      #1{\unskip}     \fi
\ifx \showLCCN     \undefined \def \showLCCN      #1{\unskip}     \fi
\ifx \shownote     \undefined \def \shownote      #1{#1}          \fi
\ifx \showarticletitle \undefined \def \showarticletitle #1{#1}   \fi
\ifx \showURL      \undefined \def \showURL       {\relax}        \fi
% invisible to TeX
\providecommand\bibfield[2]{#2}
\providecommand\bibinfo[2]{#2}
\providecommand\natexlab[1]{#1}
\providecommand\showeprint[2][]{arXiv:#2}

\bibitem[Chi et~al\mbox{.}(2020)]%
        {chi2020connections}
\bibfield{author}{\bibinfo{person}{Yu Chi}, \bibinfo{person}{Daqing He}, \bibinfo{person}{Fanghui Xiao}, {and} \bibinfo{person}{Ning Zou}.} \bibinfo{year}{2020}\natexlab{}.
\newblock \showarticletitle{Connections and Disconnections between Online Health Information Seeking and Offline Consequences}. In \bibinfo{booktitle}{\emph{Proceedings of the 14th EAI International Conference on Pervasive Computing Technologies for Healthcare}}. \bibinfo{pages}{73--84}.
\newblock


\bibitem[De~Choudhury and De(2014)]%
        {de2014mental}
\bibfield{author}{\bibinfo{person}{Munmun De~Choudhury} {and} \bibinfo{person}{Sushovan De}.} \bibinfo{year}{2014}\natexlab{}.
\newblock \showarticletitle{Mental health discourse on reddit: Self-disclosure, social support, and anonymity}. In \bibinfo{booktitle}{\emph{Proceedings of the International AAAI Conference on Web and Social Media}}, Vol.~\bibinfo{volume}{8}.
\newblock


\bibitem[Fox and Duggan(2013)]%
        {pewresearch}
\bibfield{author}{\bibinfo{person}{Susannah Fox} {and} \bibinfo{person}{Maeve Duggan}.} \bibinfo{year}{2013}\natexlab{}.
\newblock \showarticletitle{Health Online 2013}. In \bibinfo{booktitle}{\emph{Proceedings of the ninth symposium on usable privacy and security}}. \bibinfo{publisher}{Pew Research Center}, \bibinfo{pages}{1}.
\newblock


\bibitem[Kwakernaak et~al\mbox{.}(2019)]%
        {kwakernaak2019patients}
\bibfield{author}{\bibinfo{person}{Joyce Kwakernaak}, \bibinfo{person}{Just~AH Eekhof}, \bibinfo{person}{Margot~WM De~Waal}, \bibinfo{person}{Elisabeth~AM Barenbrug}, {and} \bibinfo{person}{Niels~H Chavannes}.} \bibinfo{year}{2019}\natexlab{}.
\newblock \showarticletitle{Patients’ use of the internet to find reliable medical information about minor ailments: vignette-based experimental study}.
\newblock \bibinfo{journal}{\emph{Journal of medical Internet research}} \bibinfo{volume}{21}, \bibinfo{number}{11} (\bibinfo{year}{2019}), \bibinfo{pages}{e12278}.
\newblock


\bibitem[Leon et~al\mbox{.}(2013)]%
        {leon2013matters}
\bibfield{author}{\bibinfo{person}{Pedro~Giovanni Leon}, \bibinfo{person}{Blase Ur}, \bibinfo{person}{Yang Wang}, \bibinfo{person}{Manya Sleeper}, \bibinfo{person}{Rebecca Balebako}, \bibinfo{person}{Richard Shay}, \bibinfo{person}{Lujo Bauer}, \bibinfo{person}{Mihai Christodorescu}, {and} \bibinfo{person}{Lorrie~Faith Cranor}.} \bibinfo{year}{2013}\natexlab{}.
\newblock \showarticletitle{What matters to users? Factors that affect users' willingness to share information with online advertisers}. In \bibinfo{booktitle}{\emph{Proceedings of the ninth symposium on usable privacy and security}}. \bibinfo{pages}{1--12}.
\newblock


\bibitem[Liao(2019)]%
        {liao2019sharing}
\bibfield{author}{\bibinfo{person}{Yuting Liao}.} \bibinfo{year}{2019}\natexlab{}.
\newblock \showarticletitle{Sharing Personal Health Information on Social Media: Balancing Self-presentation and Privacy}. In \bibinfo{booktitle}{\emph{Proceedings of the 10th International Conference on Social Media and Society}}. \bibinfo{pages}{194--204}.
\newblock


\bibitem[McCarthy et~al\mbox{.}(2017)]%
        {mccarthy2017did}
\bibfield{author}{\bibinfo{person}{Danielle~M McCarthy}, \bibinfo{person}{Grant~N Scott}, \bibinfo{person}{D~Mark Courtney}, \bibinfo{person}{Alyssa Czerniak}, \bibinfo{person}{Amer~Z Aldeen}, \bibinfo{person}{Stephanie Gravenor}, {and} \bibinfo{person}{Scott~M Dresden}.} \bibinfo{year}{2017}\natexlab{}.
\newblock \showarticletitle{What did you Google? Describing online health information search patterns of ED patients and their relationship with final diagnoses}.
\newblock \bibinfo{journal}{\emph{Western Journal of Emergency Medicine}} \bibinfo{volume}{18}, \bibinfo{number}{5} (\bibinfo{year}{2017}), \bibinfo{pages}{928}.
\newblock


\bibitem[Schoenherr and White(2014)]%
        {schoenherr2014interactions}
\bibfield{author}{\bibinfo{person}{Georg~P Schoenherr} {and} \bibinfo{person}{Ryen~W White}.} \bibinfo{year}{2014}\natexlab{}.
\newblock \showarticletitle{Interactions between health searchers and search engines}. In \bibinfo{booktitle}{\emph{Proceedings of the 37th international ACM SIGIR conference on Research \& development in information retrieval}}. \bibinfo{pages}{143--152}.
\newblock


\bibitem[White and Horvitz(2009)]%
        {white2009cyberchondria}
\bibfield{author}{\bibinfo{person}{Ryen~W White} {and} \bibinfo{person}{Eric Horvitz}.} \bibinfo{year}{2009}\natexlab{}.
\newblock \showarticletitle{Cyberchondria: studies of the escalation of medical concerns in web search}.
\newblock \bibinfo{journal}{\emph{ACM Transactions on Information Systems (TOIS)}} \bibinfo{volume}{27}, \bibinfo{number}{4} (\bibinfo{year}{2009}), \bibinfo{pages}{1--37}.
\newblock


\end{thebibliography}

\appendix
\section{Appendix: List of Subreddits}
\hfill \break
\begin{tabular}{ l l }
 1. r/Medical & 6. r/PublicHealth  \\ 
 2. r/MedicineCommunity & 7. r/AskATherapist \\  
 3. r/Healthcare & 8. r/ItsNeverLupus \\
 4. r/Telemedicine & 9. r/StackAdvice \\
 5. r/BehaviorAnalysis & 10. r/MentalHealth
\end{tabular}

\end{document}